\documentstyle[amssymb,amscd,12pt]{amsart}

\theoremstyle{definition}

\theoremstyle{definition}

\theoremstyle{remark}

\begin{document}

\newcommand{\thmref}[1]{Theorem~\ref{#1}}
\newcommand{\secref}[1]{\S~\ref{#1}}
\newcommand{\lemref}[1]{Lemma~\ref{#1}}
\newcommand{\propref}[1]{Proposition~\ref{#1}}
\newcommand{\corref}[1]{Corollary~\ref{#1}}
\newcommand{\remref}[1]{Remark~\ref{#1}}
\newcommand{\nc}{\newcommand}
\nc{\on}{\operatorname}
\nc{\ch}{\mbox{ch}}
\nc{\Z}{{\Bbb Z}}
\nc{\C}{{\Bbb C}}
\nc{\cond}{|\,}
\nc{\bib}{\bibitem}
\nc{\pone}{\C{\Bbb P}^1}
\nc{\pa}{\partial}
\nc{\HH}{{\cal H}}
\nc{\arr}{\rightarrow}
\nc{\larr}{\longrightarrow}
\nc{\al}{\alpha}
\nc{\ri}{\rangle}
\nc{\lef}{\langle}
\nc{\W}{{\cal W}}
\nc{\gam}{\bar{\gamma}}
\nc{\Q}{\bar{Q}}
\nc{\q}{\widetilde{Q}}
\nc{\la}{\lambda}
\nc{\ep}{\epsilon}
\nc{\su}{\widehat{\goth{s}\goth{l}}_2}
\nc{\gb}{\bar{\goth{g}}}
\nc{\g}{\goth{g}}
\nc{\hh}{\widehat{\goth{h}}}
\nc{\h}{\goth{h}}
\nc{\n}{\goth{n}}
\nc{\ab}{\widehat{\goth{a}}}
\nc{\f}{\widehat{{\cal F}}}
\nc{\is}{{\bold i}}
\nc{\V}{{\cal V}}
\nc{\M}{\widetilde{M}}
\nc{\js}{{\bold j}}
\nc{\bi}{\bibitem}
\nc{\laa}{\bar{\lambda}}
\nc{\fl}{B_-\backslash G}
\nc{\De}{\Delta}
\nc{\G}{\widehat{\goth{g}}}
\nc{\Li}{{\cal L}}
\nc{\fp}[2]{\frac{\pa}{\pa u_{#1}^{(#2)}}}
\nc{\Ve}{{\cal Vect}}
\nc{\La}{\Lambda}
\nc{\Ga}{\Gamma}
\nc{\one}{{\bold 1}}
\nc{\N}{\widehat{\n}}
\nc{\we}{\widetilde{\epsilon}}
\nc{\di}{{\cal D}}
\nc{\gl}{\goth{g}\goth{l}}
\nc{\lo}{\on{loc}}
\nc{\sw}{\goth{s}\goth{l}}
\nc{\zn}{{\bold z}}
\nc{\F}{{\cal F}}
\nc{\hb}{\nu}
\nc{\vecl}{\on{Vect} \widetilde{{\cal U}}_{\lo}}

\title{Free field realizations in representation theory and conformal field
theory}\thanks{Invited lecture at the International Congress of
Mathematicians, Z\"{u}rich, August 3-11, 1994}
\author{Edward Frenkel}
\address{Department of Mathematics, Harvard University,
Cambridge, MA 02138, USA}

\maketitle

\medskip

\noindent
Free field realizations are realizations of conformal algebras in terms of
infinite-dimensional Heisenberg or Clifford algebras. From the physics
point of view, this gives a representation of a two-dimensional conformal
field theory via a free bosonic or free fermionic theory. Mathematically,
this can be considered as ``abelianization'' of a complicated symmetry
algebra. In recent years free field realizations have been found for a
variety of conformal (super) algebras and they provided valuable insights
on representation theory and quantum field theory associated to them.

In this report, which is based mainly on my joint works with Boris Feigin,
we will focus on free field realizations of affine Kac-Moody algebras and
$\W$--algebras. We will give two constructions of free field realizations:
geometric and hamiltonian, and discuss their applications.

\section{Preliminaries}

\subsection{Heisenberg algebras.}    \label{heis}
We first introduce two types of Heisenberg algebras, $\ab$ and $\HH_S$.

Let $\goth{a}$ be a finite-dimensional linear space with a scalar product
$( \cdot,\cdot )$. We choose a basis $v_i, i=1,\ldots,N$, of $\goth{a}$. The
Heisenberg Lie algebra $\ab$ has generators $b_i(n), i=1,\ldots,N, n \in
\Z$, and $\one$, with the commutation relations $$[b_i(n),b_j(m)] = n
(v_i,v_j) \one, \quad \quad [\one,b_i(n)] = 0.$$

The second Heisenberg algebra, which we denote by $\HH_S$, where $S$
is a set, has generators $a_\al(n), a^*_\al(n), \al \in S, n \in \Z$, and
$\one$, with the commutation relations $$[a_\al(n),a_\beta^*(m)] =
\delta_{\al,\beta}
\delta_{n,-m} \one, \quad \quad [a_\al(n),a_\beta(m)] = 0, \quad
\quad [a_\al^*(n),a_\beta^*(m)] = 0,$$ and $\one$ commutes with everything.

In physics terminology, $\ab$ is the Heisenberg algebra of $N$ scalar
fields, and $\HH_S$ is a $\beta\gamma$--system. For more details, cf.
\cite{ff:weil}.

\subsection{Fock representations.}    \label{fock}
Let $\la$ be an element of $\goth{a}^*$, the dual space to $\goth{a}$, and
$\hb \neq 0$ be a complex number. We define the Fock space representation
$\pi^\hb_\la$ of $\ab$ as a module freely generated by $b_i(n),
i=1,\ldots,N, n<0$, from a vector $v_\la$, such that $b_i(n) v_\la = 0,
n>0; b_i(0) v_\la = \la(v_i) v_\la$; and on which the central element
$\one$ acts as $\hb$ times the identity. We put $\pi_\la =
\pi^1_\la$.

We also define a Fock representation $M$ of the Lie algebra $\HH_S$ as a
module freely generated by $a_\al(n), \al \in S, n<0$, and $a_\al^*(n), \al
\in S, n\leq 0$, from a vector $v$, such that
$a_\al(n) v = 0, n \geq 0; a_\al^*(n) v = 0, n>0$.
The central element $\one$ acts on $M$ as the identity.

We introduce $\Z$--gradings on the Lie algebras $\ab$ and $\HH_S$ by
putting $\deg b_i(n) = \deg a_\al(n) = \deg a_\al^*(n) = -n, \deg \one =
0$. The modules $\pi^\hb_\la$ and $M$ inherit these gradings. These modules
are always irreducible. They can be realized in a natural way in spaces of
polynomials in infinitely many variables.

\subsection{Vertex operator algebra structure.}    \label{voa}
The modules $\pi^\hb_0$ and $M$, which we call {\em vacuum modules}, carry
vertex operator algebra (VOA) structures. Recall \cite{B,FLM} that a VOA
structure is essentially a linear operation on a $\Z$--graded linear space
$V$, which associates to any homogeneous vector $A \in V$, a formal power
series, called a {\em current}, $Y(A,z) =
\sum_{m\in\Z} A_m z^m$, where $A_m: V \arr V$ is a linear operator of
degree $\deg A + m$. These series satisfy certain axioms, cf. \cite{B,FLM}.

The VOA structure on $\pi^\hb_0$ and $M$ can be described explicitly. We
will give here an explicit formula for $Y(\cdot,z)$ in the VOA $M$; the
case of $\pi^\hb_0$ was treated in detail in \cite{ff:im}, \S~4.1.

Monomials $a_{\al_1}(m_1) \ldots a_{\al_k}(m_k) a_{\beta_1}^*(n_1) \ldots
a^*_{\beta_l}(n_l) v, m_p < 0, n_q\leq 0$, form a linear basis in
$M$. The series $Y(\cdot,z)$ associated to this monomial is given by
$$C \, \,:\pa_z^{-m_1-1} a_{\al_1}(z) \ldots \pa_z^{-m_k-1} a_{\al_k}(z)
\pa_z^{-n_1} a_{\beta_1}^*(z) \ldots \pa_z^{-n_l} a_{\beta_l}^*(z):,$$
where $C = [(-m_1-1)! \ldots (-m_1-1)! (-n_1)! \ldots (-n_l)!]^{-1}$, and
\begin{equation}    \label{betgam}
a_i(z) = \sum_{n \in \Z} a_i(n) z^{-n-1}, \quad \quad a_j^*(z) =
\sum_{n\in\Z} a_j^*(n) z^{-n}.
\end{equation}

The Fourier coefficients of currents form a Lie algebra, which lies in a
certain completion of the universal enveloping algebra $U(\ab)$ or
$U(\HH_S)$ factored by the ideal generated by $(\one - \hb)$ or $(\one -
1)$, respectively. Following \cite{ff:gd}, we call this Lie algebra {\em
local completion} of the universal enveloping algebra and denote it by
$U_\hb(\ab)_{\on{loc}}$ or $U(\HH_S)_{\on{loc}}$, respectively.  We also
put $U(\ab)_{\lo} = U_1(\ab)_{\lo}$.

Let us also define {\em bosonic vertex operators} $V^\nu_\gamma(z) =
\sum_{n\in\Z} V^\nu_\gamma(n) z^{-n+(\gamma,\la)} =$
\begin{equation}    \label{vertex}
= T_\gamma z^{(\gamma,\la)}
\exp \left( -\sum_{n<0} \frac{\gamma(n) z^{-n}}{n} \right) \exp \left(
-\sum_{n>0} \frac{\gamma(n) z^{-n}}{n} \right) ,
\end{equation}
where $\gamma \in \goth{a}^* \simeq \goth{a}$ and $T^\nu_\gamma:
\pi^\nu_\la \arr \pi^\nu_{\la+\gamma}$ is such that $T_\gamma \cdot v_\la =
v_{\la+\gamma}$ and $[T^\nu_\gamma,b_i(n)]=0, n<0$. Thus, $V^\nu_\gamma(n),
n \in \Z$, are well-defined linear operators acting from $\pi^\nu_\la$ to
$\pi^\nu_{\la+\gamma}$. They appear naturally in the context of VOA
\cite{ff:im}, \S~4.2.

\subsection{Affine algebras.}    \label{aff}
Let $\g$ be a finite-dimensional simple Lie algebra over $\C$ of rank
$l$. Recall that the {\em affine Lie algebra} associated to $\g$ is the
universal central extension $\G$ of the loop algebra $L\g = \g \otimes
\C[t,t^{-1}]$. We denote by $K$ the central element of
$\G$, and for $A \in \g, n \in \Z$, we denote by $A(n)$ the element $A
\otimes t^n \in \G$.

Let $\g = \n_+ \oplus \h \oplus \n_-$ be the Cartan decomposition of $\g$.
The Lie algebra $\G$ has a {\em Cartan decomposition}: $\G =
\widetilde{\n}_+ \oplus \widetilde{\h} \oplus \widetilde{\n}_-$, where
$\widetilde{n}_\pm = \n_\pm \otimes \C 1 \oplus \g \otimes t^{\pm 1}
\C[t^{\pm 1}]$, and $\widetilde{\h} = \h \otimes \C 1 \oplus \C K$.
We denote by $e_i, h_i, f_i, i=0,\ldots,l$, the standard generators of $\G$
\cite{K}.

The Lie algebra $\G$ also has a {\em loop decomposition}: $\G =
\widehat{\n}_+ \oplus \widehat{\h} \oplus \widehat{\n}_-$, where $\N_\pm =
\n_\pm \otimes \C[t,t^{-1}]$, and $\hh = \h \otimes \C[t,t^{-1}] \oplus \C K$.
Note that the latter subalgebra is the Heisenberg Lie algebra associated
with $\h$, where the scalar product is the restriction of the Killing
form, and $K$ plays the role of $\one$.

There is a family of VOAs associated to $\G$, the {\em vacuum
representations of level} $k$, $k \in \C$: $$V_k = U(\G)
\otimes_{U(\g \otimes \C[t]
\oplus \C K)} \C_k,$$ where $\C_k$ stands for the trivial one-dimensional
representation of the Lie subalgebra $\g \otimes \C[t]$ of $\G$, on which
$K$ acts by multiplication by $k$. Its $\Z$--grading is inherited from the
standard $\Z$--grading on $\G$, such that $\deg A(n) = -n, \deg K = 0$.

The Fourier components of currents of $V_k$ form a Lie algebra
$U_k(\G)_{\on{loc}}$, which is called the local completion of the universal
enveloping algebra of $\G$. It lies in a certain topological completion of
$U(\G)/(K-k)U(\G)$, cf. \cite{ff:gd}.

\subsection{Category ${\cal O}$ and Verma modules.}    \label{verma}
Category ${\cal O}$ can be defined for an arbitrary Kac-Moody algebra using
its Cartan decomposition. It consists of modules, on which (1) the upper
nilpotent subalgebra ($\widetilde{\n}_+$ in the case of $\G$) acts locally
nilpotently, and (2) the Cartan subalgebra ($\widetilde{\h}$ in the case of
$\G$) acts semi-simply \cite{bgg,rw,dgk}. To motivate this definition in
the affine case, it is worth mentioning that the Lie group of
$\widetilde{\n}_+$ is an analogue of the compact subgroup of a simple Lie
group $G$ over a local non-archimedian field. Elements of
$\widetilde{\n}_+$ also annihilate the vacuum state of the corresponding
quantum field theory.

The fundamental objects of the category ${\cal O}$ are Verma modules. Such
a module is the induced representation $$M_\la = U(\G)
\otimes_{U(\widetilde{\n}_+ \oplus \widetilde{\h})} \C_\la,$$ where $\C_\la$
is the one-dimensional $\widetilde{\n} \oplus \widetilde{\h}$--module, on
which the first summand acts by $0$, and the second summand acts according
to its character $\la \in \widetilde{\h}^*$, which is called the {\em
highest weight}. We will write $\la = (\bar{\la},k)$, where $\bar{\la} \in
\h^*$ is the restriction of $\la$ to $\h \subset \widetilde{\h}$, and $k =
\la(K)$, $k$ is called {\em level}. All irreducible objects in ${\cal O}$
can be obtained as quotients of Verma modules.

\section{Geometric approach to free field realizations}

In this section we will give a construction of a family of free field
representations of affine algebras, which we call {\em Wakimoto
modules}. These modules were defined by Wakimoto \cite{wak} for the
simplest affine algebra $\su$ and by Feigin and the author
\cite{ff:usp} for an arbitrary affine algebra.

As was said at the beginning, our goal is to construct explicitly an
embedding of an affine algebra $\G$ into a Heisenberg algebra. A suitable
Fock representation of the latter will then provide a family of
$\G$--modules from the category ${\cal O}$, which have many nice
properties. More precisely, we will construct an embedding of $\G$ into the
local completion of the universal enveloping algebra of a Heisenberg
algebra. This embedding yields a bosonization of the
Wess-Zumino-Novikov-Witten (WZNW) model, which is a conformal field theory
associated to $\G$, and allows to compute correlation functions of this
model (see the next section).

Our geometric construction \cite{ff:usp,ff:cmp} of Wakimoto modules can be
considered as a generalization of the Borel-Weil-Bott construction of
representations of semi-simple Lie algebras, which we now recall.

\subsection{Embeddings of $\g$ into differential operators.} The universal
enveloping algebra of a Heisenberg algebra, in which the central element is
identified with $1$, is nothing but the algebra of (algebraic) differential
operators on an affine space (also called {\em Weyl algebra}). If $X$ is a
homogeneous space of the Lie group of $\g$, then $\g$ acts infinitesimally
on $X$ and hence on any open subspace of $X$ by vector fields. If we choose
an open subspace isomorphic to an affine space, we obtain an embedding of
$\g$ into a Heisenberg algebra. We can even obtain a family of such
embeddings by considering equivariant line bundles over $X$.

Let $\g$ be a simple Lie algebra, and $X$ be its flag manifold $G/B_-$,
where $G$ is the Lie group of $\g$, and $B_-$ is its Borel subgroup: the
Lie algebra of $\n_- \oplus \h$. As an open subspace of $X$, we will take
the {\em big cell} $\, \, {\cal U}$, which is the orbit of the unit coset
under the action of the Lie group $N_+$. Thus, we obtain an embedding
$\epsilon: \g \arr \on{Vect} {\cal U}$, where $\on{Vect} {\cal U}$ is the
Lie algebra of vector fields on ${\cal U}$.

The big cell ${\cal U}$ is isomorphic to the Lie group $N_+$ and hence to
the Lie algebra $\n_+$ via the exponential map. Therefore we can choose
coordinates $x_\al, \al \in \De_+$, where $\De_+$ is the set of positive
roots of $\g$, on ${\cal U}$, such that $x_\al$ has weight $\al$ with
respect to the action of the Cartan subgroup of $G$ on $N_+$.

Recall that we have a standard filtration $0 \subset \di_0 \subset \di_1
\subset \ldots$ on the algebra ${\cal D}$ of
differential operators on ${\cal U}$, where $\di_i$ is the space of
differential operators of order less than or equal to $i$. We have the
exact sequence of Lie algebras
\begin{equation}    \label{exseq}
0 \arr {\cal D}_0 \arr {\cal D}_1 \arr \on{Vect} {\cal U} \arr 0.
\end{equation}
In order to construct a map from $\g$ to ${\cal D}$, we have to lift the
map $\epsilon$ to a map $\ep': \g \arr \di_1$.

This can be done because the sequence \eqref{exseq} splits: $H^2(\on{Vect}
{\cal U},\di_0) = 0$. The inverse map $\on{Vect} {\cal U} \arr \di_1$ can
be constructed by mapping vector fields to the differential operators of
order $1$, which annihilate constants.

However, such a lifting is not unique: the space of liftings is a torsor
over $H^1(\g,\di_0) \simeq \h^*$. Thus we obtain a family of embeddings
$\ep'_\la: \g \arr \di_1 \subset \di$, parametrized by weights $\la \in
\h^*$ (to $\la=0$ we associate the embedding, whose image annihilates
constants). Now we can restrict to the image of $\ep'_\la$ the ${\cal
D}$--module $\C[{\cal U}]$ of regular functions on ${\cal U}$. Note that
this module is a Fock representation of ${\cal D}$, generated by $x_\al,
\al \in \De_+$, from a vector $v$ (a constant) satisfying: $\pa/\pa x_{\al}
\cdot v = 0, \al \in \De_+$.

\smallskip
\noindent{\bf Proposition 1} {\em The restriction of $\C[{\cal U}]$ to the
image of the homomorphism $\ep'_\la$ defines a $\g$--module, which is
isomorphic to the module $M_\la^*$ contragradient to the Verma module over
$\g$ with highest weight $\la$.}

\smallskip
\noindent{\bf Example.} The module $M_\la^*, \la \in \C$, over
$\goth{s}\goth{l}_2$ can be realized in the space $\C[x]$ as
follows: $$e = \frac{\pa}{\pa x}, \quad \quad h = -2 x \frac{\pa}{\pa x} +
\la, \quad \quad f = - x^2 \frac{\pa}{\pa x} + \la x.$$

\subsection{Semi-infinite flag manifold.}    \label{sif}
Our construction of Wakimoto modules essentially exploits the same idea: we
should find an appropriate homogeneous space of the Lie group of $\G$ and
try to embed $\G$ into the algebra of differential operators on a big
cell. We should then choose a module over this algebra in such a way that
its restriction to $\G$ lies in the category ${\cal O}$.

We can take as a homogeneous space, the standard flag manifold of
$\G$, i.e. the quotient of the Lie group of $\G$ by the standard Borel
subgroup -- the Lie group of $\widetilde{\n}_-
\oplus \widetilde{\h}$. The construction of the previous section carries
over to this case without any difficulties and gives a realization of
contragradient Verma modules over $\G$ in the space of functions on the big
cell of this flag manifold.

However, there are other possibilities, which have no analogues in the
finite-dimensional picture. The reason is that in the affine algebra there
are many different Borel subalgebras, which are not conjugated to each
other. One of them is $\N_+ \oplus \h
\otimes t\C[t]$, a Lie subalgebra of loops to the Borel subalgebra of $\g$.
To this subalgebra there corresponds the {\em semi-infinite flag manifold}
$\widetilde{X}$, which is the quotient of the loop group $LG$ by the
connected component $LB_-^0$ of the loop group of the Borel subgroup $B_-$
of $G$. One can also describe $\widetilde{X}$ as the universal covering
space of the loop space of the flag manifold $X$ of $G$, cf. \cite{ff:cmp},
\S~4.

We take as the big cell on $\widetilde{X}$, the orbit $\widetilde{{\cal
U}}$ of the unit coset under the action of the loop group of $N_+$,
$LN_+$. This orbit is isomorphic to $LN_+$, and hence to $\n_+
\otimes \C[t,t^{-1}]$, because $N_+ \simeq \n_+$ via the exponential map.
Hence we obtain coordinates $x_\al(n) = x_\al \otimes t^n, \al \in
\De_+, n\in\Z$, on the big cell $\widetilde{{\cal U}}$.

We can now identify the algebra of differential operators on
$\widetilde{{\cal U}}$ with $U(\HH_{\De_+})$, where $\HH_{\De_+}$
is the Heisenberg algebra introduced in \secref{heis}, factored by the
relation $\one=1$. Namely, we identify $a_\al^*(n)$ with $x_\al(-n)$ and
$a_\al(n)$ with $\pa/\pa x_\al(n)$.

The loop algebra $L\g$ infinitesimally acts on $\widetilde{{\cal U}}$ by
vector fields. These vector fields are actually infinite sums, and
therefore lie in a completion of the Lie algebra of vector fields on
$\widetilde{{\cal U}}$. If we could lift these vector fields to a
completion of the algebra of differential operators, we would define an
$L\g$--module from a module over the differential operators.

Of course, we could take as such a module, the space of functions on
$\widetilde{{\cal U}}$, i.e. the module generated by a vector $v$, such
that $\pa/\pa x_\al(n) \cdot v = 0, n\in\Z$. But then the resulting module
would not lie in the category ${\cal O}$, cf. \cite{jk}. In order to obtain
a module from the category ${\cal O}$, we should instead take the space of
{\em $\delta$--functions} on $\widetilde{{\cal U}}$ with support on its
subspace $\n_+ \otimes \C[t] \subset \n_+ \otimes \C[t,t^{-1}] =
\widetilde{{\cal U}}$ of ``semi-infinite dimension''. This module is
therefore generated by a vector $v$, such that $\pa/\pa x_\al(n) \cdot v
= 0, n\geq 0$, and $x_\al(n) \cdot v = 0, n<0$. As an ${\cal
H}_{\De_+}$--module, this is precisely the Fock module $M$, defined in
\secref{fock}.

Thus, we want to make $M$ into an $L\g$--module. Therefore the completion
of $U(\HH_{\De_+})$, into which we should embed $L\g$, is the local
completion $U(\HH_{\De_+})_{\on{loc}}$, defined in
\secref{voa}, because its action on $M$ is well-defined.

\subsection{Wakimoto modules.}
There is a filtration of $U(\HH_{\De_+})_{\on{loc}}$ by the powers of
the generators $a_\al(n)$: $0
\subset \di_{0,\on{loc}} \subset \di_{1,\on{loc}} \subset \ldots$. We have
the exact sequence:
\begin{equation}    \label{exseq1}
0 \arr \di_{0,\lo} \arr \di_{1,\lo} \arr \vecl \arr 0,
\end{equation}
and an embedding $\widetilde{\ep}: L\g \arr \vecl$.

In order to make $M$ into a module over the loop algebra, we have to lift
the map $\widetilde{\ep}$ to a map $\widetilde{\ep}': L\g \arr
\di_{1,\lo}$. However, this can not be done, because in contrast to the
finite-dimensional case, the exact sequence
\eqref{exseq1} {\em does not split}. Indeed, it defines a class in the
cohomology group $H^2(\vecl,\di_{0,\lo})$, which was shown to be
one-dimensional \cite{ff:cmp}, \S~5.1.

This fact can be explained as follows. The Lie algebra $U({\cal
H}_{\De_+})_{\on{loc}}$ consists of infinite sums of monomials in
$a_\al(n), a_\al^*(n)$. In order to make them act on the space $M$ we had
to regularize them by means of normal ordering, cf.
\secref{voa}. This normal ordering distorts commutation relations in such a
way that in the commutator of two elements of $\di_{1,\lo}$ there appears
an extra term lying in $\di_{0,\lo}$ (it is given by the sum of all double
contractions in the Wick formula). This extra term defines a non-trivial
extension \eqref{exseq1}. Note that we can not construct the inverse map
$\vecl \arr \di_{1,\lo}$, because there are no ``constants'', i.e.
elements annihilated by all $\pa/\pa x_\al(n)$, in $M$.

Still, we can salvage the situation: it turns out that the extension of
$L\g$ by $\di_{0,\lo}$ defined by \eqref{exseq1} is cohomologically
equivalent to its extension by $\C \subset \di_{0,\lo}$. It is possible
therefore to lift $\widetilde{\ep}$ to a map from the central extension
$\G$ of $L\g$ to $\di_{1,\lo}$.

\smallskip
\noindent{\bf Theorem 1} \cite{wak,ff:usp} (a) {\em There exists a Lie
algebra homomorphism $\G \arr \di_{1,\lo} \subset
U(\HH_{\De_+})_{\on{loc}}$, which maps $K$ to $-h^\vee$, where $h^\vee$ is
the dual Coxeter number of $\g$.}

(b) {\em The space of homomorphisms $\G \arr \di_{1,\lo}$ is a principal
homogeneous space over $\h \otimes \C((z)) dz$.}
\smallskip

The Fock representation $M$ now provides a family of {\em Wakimoto module}
over $\G$ of level $-h^\vee$, which is called the {\em critical
level}. Such a module, $W_{\chi(z)}$ is attached to an arbitrary
$H^L$--connection on the punctured formal disc, $\pa/\pa z + \chi(z)$,
where $H^L$ is the dual group of the Cartan subgroup of $G$. We see that
the category ${\cal O}$ at the critical level is much larger than at other
levels. In fact, irreducible objects of this category (which are
subquotients of Wakimoto modules) are parametrized by $G^L$--connections on
the formal punctured disc satisfying a special transversality condition.
These irreducible modules can be constructed as quotients of Verma modules
by characters of the center of the local completion of $U(\G)$, which are
parametrized precisely by such connections, cf. \cite{ff:gd} and
\secref{ds} below. Here $G^L$ is the Langlands dual group of $G$, and this
fact can be used for constructing geometric {\em Langlands correspondence}
(Drinfeld).

\smallskip
\noindent{\bf Example.} Here we write down explicit formulas for the
embedding of $\widehat{\goth{s}\goth{l}}_2$ into $U({\cal H})_{\on{loc}}$,
depending on $\chi(z) = \sum_{n\in\Z} \chi_n z^{-n-1} \in \C((z))$. We will
write $x(z) = \sum_{n\in\Z} x(n) z^{-n-1}$ for $x \in \sw_2$.  The currents
$a(z)$ and $a^*(z)$ are defined by
\eqref{betgam} (we omit unnecessary subscripts):
$$e(z) = a(z), \quad \quad h(z) = - 2 :a(z) a^*(z): + \chi(z),$$ $$f(z) = -
:a(z) a^*(z) a^*(z): + \chi(z) a^*(z) - 2
\pa_z a^*(z).$$ These formulas first appeared in \cite{wak}. Explicit
formulas for $\widehat{\sw}_n$ first appeared in \cite{ff:usp}.

It is not difficult to generalize this construction to an arbitrary level.

\smallskip
\noindent{\bf Theorem 2} \cite{wak,ff:usp} {\em There is a structure of
$\G$--module of level $k$ from the category ${\cal O}$ on $W_{\chi,k} = M
\otimes \pi^{k+h^\vee}_{\chi}$.}

\subsection{Two-sided resolution.} One can construct an analogue of the
Bernstein-Gelfand-Gelfand (BGG) resolution \cite{bgg,rw}, which consists of
Wakimoto modules. Any element $s$ of the Weyl group ${\cal W}$ of $\G$ can
be uniquely written as a product $\bar{s} \cdot \gamma$, where $\bar{s}$ is
an element of the Weyl group of $\g$, and $\gamma$ is an element of the
root lattice of $\g$. Put $lt(s) = l(\bar{s}) + 2
(\bar{\rho}^\vee,\gamma)$, where $l(\bar{s})$ is the usual length of
$\bar{s}$, and $\bar{\rho}^\vee \in \h^*$ is such that
$(\bar{\rho}^\vee,\al_i) = 1, i=1,\ldots,l$; we call $lt(w)$ the {\em
modified length} of $w$ \cite{ff:cmp}. Let $\la$ be a dominant integral
weight.

\smallskip
\noindent{\bf Theorem 3} \cite{ff:cmp} {\em There exists a complex of
$\G$--modules $R^*_\la$, where \[ R^n_\la = \oplus_{s \in {\cal W},
lt(s)=n} W_{s(\la+\rho)-\rho}, \] whose cohomology is non-zero only in
dimension $0$, where it is isomorphic to the irreducible $\G$--module
$L_\la$ of highest weight $\la$.}
\smallskip

In contrast to the usual BGG resolution of $\G$, $R^*_\la(\G)$ is
two-sided. In the case of $\widehat{\sw}_2$ these resolutions were
constructed explicitly in \cite{ff:cmp}, \S~7.3 and in
\cite{bf} (they are closely connected with
similar resolutions over the Virasoro algebra constructed in
\cite{fel}). In \cite{bmp1,bmp2} a remarkable connection between $R^*_\la$
and resolutions over the quantum group $U_q(\g)$ with $q = \exp 2\pi
i/(k+h^\vee)$ was found, and explicit formulas for the differentials of
$R^*_\la$ were proposed.

\subsection{Remarks.}
(1) The construction of Wakimoto modules is a
semi-infinite version of the construction of induced and coinduced
modules. In fact, it is not difficult to construct in a similar fashion a
$\G$--bimodule $\widetilde{U}_k(\G)$ on which $\G$ acts on the left with
level $k$ and on the right with level $-2 h^\vee-k$,
so that $W_{\chi,k} \simeq \on{Tor}_{\infty/2}^{\hh \oplus
\N_-}(\widetilde{U}_k(\G),\pi^{k+h^\vee}_\chi)$, where
$\on{Tor}_{\infty/2+*}$ is the semi-infinite Tor functor
\cite{fei} (compare with the construction of Verma modules from
\secref{verma}).

(2) By construction, the modules $W_{\chi(z)}$ and $W_{\chi,k}$ are free
over the Lie algebra $\N_+ \cap \widetilde{\n}_-$ and co-free over the Lie
algebra $\N_+ \cap \widetilde{\n}_+$. Therefore they are {\em flat} over
$\N_+$ in the sense of semi-infinite cohomology:
$H^{\infty/2+i}(\N_+,W_{\chi,k}) = \pi^{k+h^\vee}_\chi$, if $i=0$, and $0$,
if $i \neq 0$ (note that $H^i(\widetilde{\n}_+,M^*_\la) = \C_\la$, if
$i=0$, and $0$, if $i \neq 0$, where $H^i$ stands for the usual Lie algebra
cohomology functor). Using this result and the two-sided resolution
$R^*_\la$ , we computed in \cite{ff:cmp}, Theorem~4, the semi-infinite
cohomology of $\N_+$ with coefficients in $L_\la$.

(3) In \cite{ff:cmp} a more general construction is given, which associates
to an arbitrary parabolic subalgebra $\goth{p}$ of $\g$, a ``Borel
subalgebra'' of $\G$. These ``Borel subalgebras'' are not conjugated with
each other and therefore lead to different flag manifolds. Generalized
Wakimoto modules can be defined as delta-functions on these manifolds. They
are flat with respect to the corresponding ``Borel subalgebra''. In
particular, $M_{\chi,k}^*$ corresponds to $\goth{p} = \g$ and $W_{\chi,k}$
corresponds to $\goth{p} = \h \oplus \n_+$.

(4) Wakimoto modules are related to the Shubert cell decomposition of the
semi-infinite flag manifold in the same way as the Verma modules are
related to the Shubert cell decomposition of the usual flag manifold,
cf. \cite{ff:cmp}. In particular, the Floer cohomology of the semi-infinite
flag manifold is the double of the semi-infinite cohomology of the Lie
algebra $\N_+$ (compare with the finite-dimensional case \cite{kos}).

(5) One can show that if $\chi(z)$ is of the form $\chi/z, \chi \in
\h^*$, and does not lie on certain hyperplanes, then $W_{\chi(z)}$ is
isomorphic to the irreducible quotient of the Verma module
$M_{\chi,-h^\vee}$. This gives a simple proof
\cite{ff:kni} of the Kac-Kazhdan conjecture \cite{kk} on characters of
irreducible modules.

(6) The character of the module $W_{\chi,k}$ coincides with the character
of the Verma module $M_{\chi,k}$ and for generic values of $\chi$ and $k$
they are irreducible isomorphic to each other. When they are not
irreducible, they may have different composition series, cf., e.g.,
\cite{ff:kni} and \cite{bf} in the case of $\widehat{\sw}_2$. A surprizing
fact \cite{f:det} is that if $k$ is real and {\em less} than $-h^\vee$,
then $W_{\chi,k} \simeq M^*_{\chi,k}$ for positive $\chi$.

\section{Solutions of the Knizhnik-Zamolodchikov equation.}    \label{sol}

In this section we will outline the application of the Wakimoto
realization to the computation of correlation functions (or conformal
blocks) in the WZNW model. It is known that in genus zero they satisfy a
system of PDE, which are called Knizhnik-Zamolodchikov (KZ)
equations. Wakimoto realization allows one to express these correlation
functions as integrals of much simpler correlation functions of free
bosonic fields. This gives the Schechtman-Varchenko solutions of the KZ
equations.

\subsection{Genus zero conformal blocks}    \label{conf}
Let us recall the definition of the space of conformal blocks in the WZNW
model. Consider the projective line $\pone$ with a global coordinate $t$
and $N$ distinct finite points $z_1,\ldots,z_N \in \pone$. In the
neighborhood of each point $z_i$ we have the local coordinate $t-z_i$;
denote $\widetilde{\g}(z_i) = \g \otimes
\C((t-z_i))$. Let $\G_N$ be the extension of the Lie algebra
$\oplus_{i=1}^N \widetilde{\g}(z_i)$ by one-dimensional center $\C K$, such
that its restriction to any summand $\widetilde{\g}(z_i)$ coincides with
the standard extension. The Lie algebra $\G_N$ naturally acts on the
$N$--fold tensor product of $\G$--modules $\otimes_{i=1}^N M_i$
of a given level $k \neq -h^\vee$.

Let $C_N$ be the space $\C^N$ with coordinates $z_1,\ldots,z_N$ without the
diagonals and $C'_N$ be the space $\C_N \times \pone$ with coordinates
$z_1,\ldots,z_N$ and $t$ without the diagonals. Denote by ${\cal B}({\bold
z})$ and ${\cal B}'({\bold z})$, where ${\bold z} = (z_1,\ldots,z_N)$, the
algebras of regular functions on $C_N$ and $C'_N$, respectively.

Introduce the Lie algebras $\G_N({\bold z}) = \G_N
\otimes_{\C} {\cal B}({\bold z})$, and $\g_{\bold z} = \g
\otimes_{\C} {\cal B}'({\bold z})$. By expanding elements of $\g_{\zn}$ in
Laurent power series in the local coordinates $t-z_i$ at each point $z_i$,
we obtain an embedding $\g_{\zn} \arr \G_N({\bold z})$.

The Lie algebra $\G_N({\bold z})$ acts on the space
$X(M_1,\ldots,M_N;{\bold z}) = \on{Hom}_\C(M_1
\otimes \ldots \otimes M_N,{\cal B}({\bold z}))$. Denote by
$H(M_1,\ldots,M_N;{\bold z})$ the ${\cal B}({\bold z})$--module of
$\g_{\bold z}$--invariants of
\newline
$X(M_1,\ldots,M_N;{\bold z})$. One has:
$H(M_1,\ldots,M_N;{\bold z}) \simeq H(M_1,\ldots,M_N) \otimes_\C {\cal
B}({\bold z})$, where $H(M_1,\ldots,M_N)$ is called the {\em space of
conformal blocks}.

Consider the operators $\nabla_i = \pa/\pa z_i - L_{-1}^{(i)},
i=1,\ldots,N$, on the space $X(M_1,$ $\ldots,M_N;{\bold z})$, where
$L_{-1}^{(i)}$ denotes the operator acting as the dual of the Virasoro
generator $L_{-1}$ (provided by the Sugawara construction) on the $i$th
argument of $X(M_1,\ldots,M_N;{\bold z})$. One can check, cf., e.g.,
\cite{ffr}, Lemma~4, that the operators $\nabla_i$ commute with each other
and normalize the action of the Lie algebra $\g_{{\bold z}}$. Therefore
they define a flat connection on the trivial bundle over $C_N$ with the
fiber $H(M_1,\ldots,M_N)$.

\subsection{The KZ equations.}    \label{kz}
Now let us choose as the modules $M_i$, the Wakimoto modules
$W_{\la_i,k}$. Recall that $W_{\la,k}$ is the tensor product $M \otimes
\pi^{k+h^\vee}_\la$. Consider its subspace $\C[a_\al^*(0)]_{\al \in
\De_+} v \otimes v_\la$. As a module over the constant subalgebra
$\g \subset \G$, it is isomorphic to the contragradient Verma module
$M^*_\la$. For any $\Psi({\bold z}) \in
H(W_{\la_1,k},\ldots,W_{\la_N,k};{\bold z})$ denote by $\psi({\bold z})$
its restriction to the subspace $\otimes_{i=1}^N M^*_{\la_i} \subset
\otimes_{i=1}^N W_{\la_i,k}$. Thus, $\psi({\bold z})$ can be considered as
a vector in $\otimes_{i=1}^N M_{\la_N}$.

\smallskip
\noindent{\bf Lemma 1} {\em If $\nabla_i \Psi({\bold z}) = 0$ for
$i=1,\ldots,N$, then $\psi({\bold z})$ satisfies the system of equations
\begin{equation}    \label{kze}
(k+h^\vee) \frac{\pa \psi({\bold z})}{\pa z_i} = H_i \cdot \psi({\bold
z}), \quad \quad i=1,\ldots,N,
\end{equation}
where $H_i = \sum_{j\neq i} I_a^{(i)} I_a^{(j)}/(z_i-z_j)$, and
$I_a^{(i)}$ denotes an element of an orthonormal basis $\{ I_a \}$ of $\g$
acting on the $i$th factor of $\otimes_{i=1}^N M_{\la_i}$.}
\smallskip

For the proof of Lemma 1, cf., e.g., \cite{ffr}, \S~6. The equations
\eqref{kze} are called the {\em KZ equations} \cite{kniz}.

\subsection{Bosonic correlation functions.} In the same way as in
\secref{conf} we can define spaces of conformal blocks with respect to the
Heisenberg algebra $\hh \oplus
\HH_{\De_+}$ \cite{ffr}. Denote by $J_p({\bold x})$ the space associated to
the tensor product of Wakimoto modules $\otimes_{i=1}^p W_{\chi_i,k}$,
where ${\bold x} = (x_1,\ldots,x_p)$ \cite{ffr}, \S~6. This is a free
${\cal B}({\bold x})$--module with one generator $\vartheta_p$, such that
$\vartheta_p({\bold v}_p) = 1$, where ${\bold v}_p$ is the tensor product
of the highest weight vectors of $W_{\chi_i,k}, i=1,\ldots,p$.

In the same way as in the previous section one can show that the operators
$\nabla_i = \pa/\pa x_i - L_{-1}^{(i)}, i=1,\ldots,p$, act on the space
$J_p({\bold x})$. If $\Phi \in J_p({\bold x})$ is such that $\nabla_i \Phi
= 0$, then $\varphi = \Phi({\bold v}_p) \in {\cal B}({\bold x})$ satisfies
the system of equations
\begin{equation}    \label{bosonkz}
(k+h^\vee) \frac{\pa \varphi}{\pa x_i} =
\sum_{j \neq i} \frac{(\chi_i,\chi_j)}{x_i-x_j} \varphi, \quad
\quad i=1,\ldots,p.
\end{equation}
These equations are the analogues of the KZ equations
for the Heisenberg algebra, but they are much simpler: the unique up to a
constant factor solution is $\varphi_p = \prod_{i<j}
(x_i-x_j)^{(\chi_i,\chi_j)/(k+h^\vee)}$. Denote $\tau_p =
\varphi_p \cdot \vartheta_p \in J_p({\bold x})$. This is the
correlation function of the scalar bosonic filed.

\subsection{Solutions.}
Now put $p=N+m$, $x_i = z_i, \chi_i = \la_i, i=1,\ldots, N$, and $x_{N+j} =
w_j, \chi_{N+j} = -\al_{i_j}, j=1,\ldots,m$. Then
$\varphi_{N,m} =$ $$ \prod_{i<j}
(z_i-z_j)^{(\la_i,\la_j)/(k+h^\vee)}
\prod_{i,j} (z_i-w_j)^{-(\la_i,\al_{i_j})/(k+h^\vee)}
\prod_{s<j} (w_s-w_j)^{(\al_{i_s},\al_{i_j})/(k+h^\vee)}.$$

Denote by $C_{m,\zn}$ the space $\C^m$ with coordinates $w_1,\ldots,w_m$
without all diagonals $w_j=w_i$ and all hyperplanes of the form
$w_j=z_i$. The multi-valued function $\varphi_{N,m}$ defines
a one-dimensional local system ${\cal L}$ on the space $C_{m,\zn}$.

We can define another action of the Lie algebra $\N_+$ on the module
$W_{\chi,k}$, which commutes with the action defined by the Wakimoto
realization. It comes from the {\rm right} action of the Lie algebra $\N_+$
on the big cell $\widetilde{{\cal U}}$ of the semi-infinite flag manifold,
\cite{ff:kni,ff:lmp}. We denote the operator of the right action of
$e_i(n) \in \N_+$ on $W_{\chi,k}$ by $e_i^R(n)$.

The restriction of $\tau_{N,m} = \varphi_{N,m} \vartheta_{N,m} \in
J_{N,m}({\bold z},{\bold w})$ to the subspace $$\otimes_{i=1}^N W_{\la_i}
\otimes e^R_{i_1}(-1) v_{-\al_{i_1}} \otimes \ldots \otimes
e^R_{i_m}(-1) v_{-\al_{i_m}}$$ defines an element of
$X(W_{\la_1,k},\ldots,W_{\la_N,k};{\bold z})$ depending on
$w_1,\ldots,w_m$. We denote this restriction by $\Phi_{N,m}$. For any
$\omega \in W_{\la_1,k} \otimes \ldots \otimes W_{\la_N}$, the multi-valued
form $\Phi_{N,m}(\omega) \, dw_1 \ldots d w_m$ can be considered as an
$m$--form on $C_{m,\zn}$ with coefficients in the local system ${\cal
L}$. It can be integrated over an $m$--cycle $\Delta$ with coefficients in
the dual local system ${\cal L}^*$, cf. \cite{sv:inv,sv:ann}.

Special ``screening'' properties of the vectors $e_i^R(-1) v_{-\al_i} \in
W_{-\al_i,k}$ \cite{ffr}, \S~7, make them ``invisible'' for the affine
algebra after integration and lead to the following result.

\smallskip
\noindent{\bf Theorem 4} $\int_\Delta \Phi_{N,m} \, dw_1 \ldots dw_m$ {\em
lies in} $H(W_{\la_1,k},\ldots,W_{\la_N,k};{\bold z})$ {\em and} $\nabla_i
\int_\Delta \Phi_{N,m}$ $dw_1 \ldots dw_m = 0$ {\em for} $i=1,\ldots,N$.
\smallskip

By Lemma 1 we can obtain solutions of the KZ equations by restricting this
integral to the subspace $\otimes_{i=1}^N M^*_{\la_i}$ of $\otimes_{i=1}^N
W_{\la_i,k}$. We refer the reader to \cite{aty} and
\cite{ffr} for this computation and only give the final result.

Introduce the vector $|w_1^{i_1},\ldots,w_m^{i_m}\ri \in
\otimes_{i=1}^N M_{\la_i}$ by the formula
\begin{equation}    \label{genbv}
|w_1^{i_1},\ldots,w_m^{i_m}\ri = \sum_{p=(I^1,\ldots,I^N)} \prod_{j=1}^N
\frac{f_{i^j_1}^{(j)} f_{i^j_2}^{(j)} \ldots
f_{i^j_{a_j}}^{(j)}}{(w_{i^j_1}-w_{i^j_2})(w_{i^j_2}-w_{i^j_3}) \ldots
(w_{i^j_{a_j}}-z_j)} |0\ri,
\end{equation}
where the summation is taken over all {\em ordered} partitions $I^1 \cup
I^2 \cup \ldots \cup I^N$ of the set $\{1,\ldots,m\}$, $I^j = \{
i^j_1,i^j_2,\ldots,i^j_{a_j} \}$, $f_i^{(j)}$ denotes the generator
$f_i \in \g$ acting on the $j$th component of $\otimes_{i=1}^N M_{\la_i}$,
and $|0\ri = v_{\la_1} \otimes \ldots \otimes v_{\la_N}$.

\smallskip
\noindent{\bf Corollary} {\em Let $\Delta$ be an $m$--dimensional cycle on
$C_{m,\zn}$ with coefficients in ${\cal L}^*$. The
$\otimes_{i=1}^N M_{\la_i}$--valued function $$\int_\Delta
\varphi_{N,m} \, \, |w_1^{i_1},\ldots,w_m^{i_m}\ri \, \, dw_1 \ldots dw_m$$
is a solution of the KZ equation.}
\smallskip

Thus, we obtained solutions of the KZ equations in terms of generalized
hypergeometric functions using the Wakimoto modules. These solutions were
first derived by Schechtman and Varchenko by other methods \cite{sv:inv}
(cf.  also \cite{law,miwa}).

\subsection{Remarks.}
(1) One can derive these solutions in a slightly different way as matrix
elements of compositions of {\em primary fields} and {\em screening
operators}, which can be constructed explicitly
\cite{ff:kni,ff:lmp,bf,bmp1,bmp2,aty}.

(2) The results of this section mean that a complicated ${\cal D}$--module
on the space $\C^N \backslash \{ \on{diagonals} \}$ defined by the KZ
equations \eqref{kze} can be embedded into the direct image of a much
simpler ${\cal D}$--module on a larger space $\C^{N+m} \backslash \{
\on{diagonals} \}$ defined by the equations \eqref{bosonkz}. Wakimoto
realization provides a natural explanation of this remarkable fact.

(3) Using Wakimoto modules at the {\em critical level} in a similar
fashion, it was shown in \cite{ffr} that the vector \eqref{genbv} is an
eigenvector of the {\em Gaudin hamiltonians} $H_i$, if $w_j$'s satisfy a
system of {\em Bethe ansatz} equations.

\section{Free field realizations from the theory of non-linear equations}

Local integrals of motion of non-linear integrable equations form Poisson
algebras, which in many cases can be naturally embedded into larger
Heisenberg-Poisson algebras. By quantizing this embedding we can obtain an
embedding of the algebra of quantum integrals of motion into a Heisenberg
algebra. This provides another source for free field realizations. We will
describe this construction in the case of Toda equations.

\subsection{Classical Toda field theory.}    \label{to}
Let $\g$ be a simple Lie algebra and $\al_1,\ldots,\al_l \in \h^*$ be the
set of simple roots of $\g$. The Toda equation associated to $\g$ reads
\begin{equation}    \label{toda}
\pa_\tau \pa_t \phi_i(t,\tau) = \sum_{j=1}^l (\al_i,\al_j)
\exp[-\phi_j(t,\tau)], \quad \quad i=1,\ldots,l,
\end{equation}
where each $\phi_i(t,\tau)$ is a family of functions on the circle with a
coordinate $t$, depending on the time variable $\tau$.

We will now review the hamiltonian formalism of the Toda equations
following our papers \cite{ff:im,ff:kdv}, which in turn used technique
developed in \cite{gd,kw}.

Let $\pi_0 = \C[u_i^{(n)}]_{1\leq i\leq l,m\geq 0}$ be the algebra of {\em
differential polynomials} in $u_1,\ldots,$ $u_l$, where $u_i \equiv
u_i^{(0)}$. It is $\Z$--graded according to $\deg u_i^{(n)} = n+1$, and
there is a derivation $\pa$ on $\pi_0$, such that $\pa u^{(n)} =
u^{(n+1)}$. Let us formally introduce variables $\phi_i, i =
1,\ldots,l$. For any element $\la =
\sum_{1\leq i\leq l} \la_i
\al_i$ of the root lattice of $\g$, define the linear space $\pi_\la
= \pi_0
\otimes e^{\laa}$, where $\bar{\la} = \sum_{0\leq i\leq l}
\la_i \phi_i$. We extend the action of the operator $\pa$ to $\pi_\la$ by
putting $\pa e^{\laa} = \sum_{0\leq i\leq l} \la_i u_i e^{\laa}.$
In other words, we put $\pa \phi_i = u_i$.

Denote by $\F_\la$ the quotient of $\pi_\la \otimes \C[t,t^{-1}]$ by the
subspace of total derivatives (and constants, if $\la=0$), where the action
of $\pa$ on $\pi_\la
\otimes \C[t,t^{-1}]$ is given by $\pa \otimes 1 + 1
\otimes \pa_t.$ We denote by $\int$ the projection $\pi_\la \arr \F_\la$.
The space $\F_\la$ can be viewed as the space of functionals in
$u_1(t),\ldots,u_l(t) \in \h \otimes \C[t,t^{-1}]$ of the form $\int
P(u(t),\pa_tu(t),\ldots;t) e^{\laa}(t) dt$, where $P$ is a differential
polynomial. We call $\F_0$ the {\em space of local functionals} in
$u_1(t),\ldots,u_l(t)$.

There is a unique partial Poisson bracket $\{ \cdot,\cdot \}: \F_0 \times
\F_\la \arr \F_\la$, such that: $$\{ \int u_i t^n,\int u_j t^m \} = n
(\al_i,\al_j) \delta_{n,-m}, \quad \quad \{ \int u_i t^n,\int e^{\laa} t^m
\} = (\al_i,\la) \int e^{\laa} t^{n+m},$$ cf. \cite{ff:im}, \S~2.2. The
restriction of this bracket to $\F_0$ makes it into a Lie algebra.

The equation \eqref{toda} can be presented in the hamiltonian form as
$\pa_\tau u_i(t,\tau) = \{ H,u_i(t,\tau) \},$ $i=1,\ldots,l$, where $H =
\sum_{i=1}^l \int e^{-\phi_i}$. This motivates the definition of the space
$I(\g)$ of {\em local integrals of motion} of the Toda theory associated to
$\g$ as the intersection of kernels of the operators $\Q_i = \{
\int e^{-\phi_i},\cdot \}: \F_0 \arr \F_{-\al_i}, i=1,\ldots,l$. The
bracket $\{ \cdot,\cdot \}$ satisfies the Jacobi identity, therefore
$I(\g)$ is a Lie algebra.

\subsection{Hidden nilpotent action.}    \label{nil}
Introduce linear operators $$\q_i = \sum_{1\leq i\leq l;n\geq 0}
(\al_i,\al_j) \pa^n e^{-\phi_i} \frac{\pa}{\pa u_j^{(n)}}, \quad \quad
i=1,\ldots,l,$$ acting from $\pi_0$ to $\pi_{-\al_i}$. Put $Q_i =
e^{\phi_i} \q_i: \pi_0 \arr \pi_0, i=1,\ldots,l$. The following crucial
statement was proved in \cite{ff:im}, (2.2.8).

\smallskip
\noindent{\bf Lemma 2}
{\em The operators $Q_i, i=1,\ldots,l$, and $\q_i, i=1,\ldots,l$, generate
the nilpotent Lie subalgebra $\n_+$ of $\g$. The operators $\q_i$ commute
with $\pa$ and the corresponding operators $\F_0 \arr \F_{\al_i}$ coincide
with $\Q_i$.}
\smallskip

Thus we obtain an action of $\n_+$ on $\pi_0$. Using this action, we can
compute $I(\g)$. By definition, $I(\g)$ is the $0$th cohomology of the
complex $\F_0 \stackrel{\oplus \Q_i}{\larr}
\oplus_{i=1}^l \F_{-\al_i}$. We will extend this complex further to
the right using the BGG resolution of $\g$ and then reduce the computation
to the cohomology of $\n_+$ with coefficients in $\pi_0$, cf. \cite{ff:im},
\S\S~2.3-2.4.

Recall \cite{bgg} that the BGG resolution of the trivial $\g$--module is a
complex $B_*(g)$, such that $B_j(g) =
\oplus_{l(s)=j} M_{w(\rho)-\rho}$, where $M_\la$
denotes the Verma module $\g$ with highest weight $\lambda$ and $w$ runs
over the Weyl group of $\g$.  The differential $d_j: B^q_j(g)
\rightarrow B^q_{j-1}(g)$ is an alternating sum of embeddings of Verma
modules $i_{s,s'}: M_{s'(\rho)-\rho} \rightarrow M_{s(\rho)-\rho}$, where
$l(s)=j-1, l(s')=j$ and $s \preceq s'$. Under this embedding, the highest
weight vector ${\bold 1}_{s'(\rho)-\rho}$ of $M_{s'(\rho)-\rho}$ maps to a
unique singular vector $P_{s,s'} {\bold 1}_{s(\rho)-\rho} \in
M_{s(\rho)-\rho}$, where $P_{s,s'}$ is an element of $U(\n_-)$.

We now define a complex $F^*(\g)$, such that $F^j(\g) = \oplus_{l(s) = j}
\pi_{s(\rho)-\rho}$. Let $P_{s',s}(Q):
\pi_{\rho-s(\rho)} \arr \pi_{\rho-s'(\rho)}$ be the map, obtained by
inserting into $P_{s',s} \in U(\n_-)$ the operators $\q_i$ instead of
$f_i$. Introduce the differential $\delta^j: F^{j-1}(\g) \arr F^j(\g)$ of
our complex as the alternating sum of the appropriate
$P_{s',s}(Q)$'s. Since $\q_i$'s generate $\n_+$, this differential is
nilpotent.

Moreover, this differential commutes with the action of $\pa$ on
$\pi_{s(\rho)-\rho}$ \cite{ff:im}, (2.4.9). Therefore we can define a new
complex $\F^*(\g)$ as the quotient of $F^*(\g) \otimes \C[t,t^{-1}]$ by the
total derivatives and constants. We have $\F^j(\g) = \oplus_{l(s) = j}
\F_{s(\rho)-\rho}$. By definition, $I(\g)$ is the $0$th cohomology of
$\F^*(\g)$.

\smallskip
\noindent{\bf Theorem 5} \cite{ff:im} (a) {\em  There exist elements
$W_i \in \pi_0$ of degrees $d_i+1, i=1,\ldots,l$, where $d_i$ is the $i$th
exponent of $\g$, such that the $0$th cohomology $\W(\g)$ of the complex
$F^*(\g)$ is isomorphic to the algebra of differential polynomials in
$W_1,\ldots,W_l, \W(\g) = \C[W_i^{(n_i)}]_{i=1,\ldots,l;n\geq 0}$. All
higher cohomologies of the complex $F^*(\g)$ vanish.}

(b) {\em The space $I(\g)$ is isomorphic to the quotient of $\W(\g) \otimes
\C[t,t^{-1}]$ by the total derivatives and constants, i.e. the space of
local functionals in $W_1(t),\ldots,W_l(t)$.}

\smallskip
\noindent{\bf Example} $\W(\sw_2) = \C[W^{(m)}]_{m\geq 0}$, where
$W = \frac{1}{2} u^2 - \pa u$. Thus, for $\g=\sw_2$ integrals of motion are
local functionals in $W(t) = \frac{1}{2} u(t)^2 - \pa_t u(t)$.

Let ${\cal L}$ be the Virasoro algebra, which is the central extension of
the Lie algebra $\C[t,t^{-1}] \pa_t$ of vector fields on the circle. We can
consider $W(t)$ as an element of the hyperplane in its dual ${\cal L}^*$,
which consists of the linear functionals taking value $1$ on the central
element. This hyperplane is equipped with a canonical Kirillov-Kostant
Poisson bracket. The space of local functionals on this hyperplane is
isomorphic to $I(\sw_2)$, cf. \cite{ff:im}, \S~2.1, for more details.

\smallskip
\noindent{\bf Remark} The Lie algebra $I(\g)$ is called the {\em classical
$\W$--algebra} associated to $\g$. It can be identified with the Poisson
algebra of local functionals on an infinite-dimensional hamiltonian space
obtained from the dual space to $\G$ by the {\em Drinfeld-Sokolov
reduction} \cite{ds}. For $\g=\sw_N$, it was first defined by Adler and
Gelfand-Dickey.

\subsection{Quantum integrals of motion.}    \label{quan}
In the previous section we obtained the space of integrals of motion of a
Toda theory as a Lie subalgebra of $\F_0$, which lies in the kernel of the
operators $\Q_i$. Now we want to quantize this embedding. In order to do
that, we have to quantize the Lie algebra $\F_0$ and the operators $\Q_i$.

The Lie algebra $\F_0^\hb := U_\hb(\hh)_{\lo}$ defined in \secref{voa} is a
quantum deformation of the Lie algebra $\F_0$, in the following sense. We
have a natural map $\F^\hb_0 \arr \F_0$, which sends a Fourier component of
a current, $\int :P(h_i(z),\pa_z h_i(z),\ldots): z^m dz \in F^\hb_0$ to the
local functional $\int P(u_i(t),\pa_t u_i(t),\ldots) t^m dt \in \F_0$ and
sets $\hb$ to $0$\footnote{to be more precise, we should consider $\hb$ as
a formal parameter and $\F_\la^\hb, \pi_\la^\hb$ as free modules over
$\C[[\hb]]$. Then setting $\nu$ to $0$ means taking the quotient by $\nu
\C[[\nu]]$}. Denote by $\bar{A}$ the image of $A \in F^\hb_0$ in $\F_0$
under this map. The commutator of any two elements of $F^\hb_0$ has the
form: $[A,B] = \hb \cdot C + \hb^2 (\ldots)$, where $\bar{C} = \{
\bar{A},\bar{B} \}$ is the bracket of $\bar{A}$ and $\bar{B}$ in $\F_0$,
cf. \cite{ff:im}, \S~4.2 ($\hb$ was denoted by $\beta^2$ in \cite{ff:im}).

By definition, the Lie algebra $\F^\hb_0$ is the quotient of $\pi_0^\hb
\otimes \C[z,z^{-1}]$ by the total derivatives and constants, where
$\pi_0^\hb$ is the VOA of the Heisenberg algebra $\hh$. Denote by
$\F^\hb_\la$ the quotient of $\pi_\la^\hb \otimes \C[z,z^{-1}]$ by the
total derivatives.  We have a map $\pi^\hb_\la \arr \pi_\la$, where
$\pi_\la$ was defined in
\secref{to}, which sends $h_{i_1}(n_1) \ldots h_{i_m}(n_m) v_\la \in
\pi_\la^\hb$ to $(-n_1-1)! \ldots (-n_m-1)! \, \, u_{i_1}^{(-n_1-1)}
\ldots  u_{i_m}^{(-n_m-1)} \otimes e^{\laa} \in \pi_\la$, and sets $\hb$ to
$0$.

The quantizatum deformation of the operator $\q_i: \pi_0 \arr \pi_{-\al_i}$
is the operator $\q^\hb_i := V^\nu_{-\al_i}(1): \pi^\hb_0 \arr
\pi^\hb_{-\al_i}$, where $V^\nu_\gamma(n)$ was defined by formula
\eqref{vertex}. Indeed, one can check, cf. \cite{ff:im}, (4.2.4), that
$\q^\hb_i = \hb \cdot \q_i +
\hb^2 (\ldots)$. Further, we can check that $\q^\hb_i$ commutes with the
derivative $\pa$. The corresponding operators $\Q^\hb_i: \F^\hb_0
\arr \F^\hb_{-\al_i}$ are quantum deformations of $\Q_i, i=1,\ldots,l$.

We can now define the space $I_\hb(\g)$ of {\em quantum integrals of
motion} of Toda theory associated to $\g$ as $$I_\hb(\g) = \bigcap_{i=1}^l
\on{Ker}_{\F^\hb_0} \Q^\hb_i.$$ One can check that $I_\hb(\g)$ is a
Lie subalgebra of $\F^\hb_0$, \cite{ff:im}, (4.2.8). Thus, we {\em define}
it through its embedding into $\F^\hb_0$, i.e. through its free field
realization.

We also define the space $\W_\hb(\g)$ as $$\W_\hb(\g) = \bigcap_{i=1}^l
\on{Ker}_{\pi^\hb_0} \q^\hb_i.$$ One can check that $\W_\hb(\g)$ is a VOA
\cite{ff:im}, (4.2.8).

\subsection{$\W$--algebras.}    \label{walg}
We will now deform the complex $F^*(\g)$. Vanishing of higher cohomologies
of $F^*(\g)$ will imply their vanishing for the deformed complex. This will
allow us to prove that the $0$th cohomology $\W_\hb(\g)$, and hence
$I_\hb(\g)$, are quantum deformations of $\W(\g)$ and $I(\g)$,
respectively.

The construction of the quantum complex is based on the fact that while the
operators $\q_i$ generate $U(\n_+)$, their quantum deformations, operators
$\q^\hb_i$, generate the quantized enveloping algebra $U_q(\n_+)$ with
$q=\exp \pi i \hb$ in a certain sense, cf. \cite{ff:im}, \S~4.5, for a
precise statement. This was discovered in \cite{bmp1}, cf. also
\cite{sv:ann}, where a more general connection between local systems on
configuration spaces and quantum groups was established.

We can then use a quantum deformation of the BGG resolution, cf.
\cite{ff:im}, \S~4.4, to construct the deformed complex $F_\hb^*(\g)$. As
a linear space, $F_\hb^j(\g) = \oplus_{l(s) = j} \pi^\hb_{s(\rho)-\rho}$. The
differentials are constructed using the operators $\q^\hb_i$, cf.
\cite{ff:im}, \S~4.5. The differential $\delta^1_\hb: \pi^\hb_0 \arr
\oplus_{i=1}^l \pi^\hb_{-\al_i}$ is given by the sum of the operators
$\q^\hb_i, i=1,\ldots,l$, so that the $0$th cohomology of $F^*_\hb(\g)$ is
$\W_\hb(\g)$.

Now we have a family of complexes $F^*_\hb(\g)$, depending on $\hb$. From
vanishing of higher cohomologies for $\hb=0$, cf. Theorem 5, we obtain the
following result.

\smallskip
\noindent{\bf Theorem 6} \cite{ff:im} (a) {\em For generic $\hb$ higher
cohomologies of the complex $F^*_\hb(\g)$ vanish. The $0$th cohomology,
$\W_\hb(\g)$, is a VOA, in which there exist elements $W^\hb_i$ of degrees
$d_i+1, i=1,\ldots,l$, where $d_i$ is the $i$th exponent of $\g$, such that
$\W_\hb(\g)$ has a linear basis of lexicographically ordered monomials in
the Fourier components $W^\hb_i(n_i), 1\leq i\leq l,n_i<-d_i,$ of the
currents $Y(W^\hb_i,z) = \sum_{n\in\Z} W^\hb_i(n) z^{-n-d_i-1}.$}

(b) {\em The Lie algebra $I_\hb(\g)$ of quantum integrals of motion of the
Toda theory associated to $\g$ consists of all Fourier components of
currents of the VOA $\W_\hb(\g)$.}
\smallskip

The Lie algebra $I_\hb(\g)$ is the $\W$--{\em algebra} associated to
$\g$. Such $\W$--algebras are, along with affine algebras, the main
examples of algebras of symmetries of conformal field theories
\cite{bs}. Part (a) of Theorem 6 means that its VOA $\W_\hb(\g)$ is ``freely
generated'' by the currents $Y(W^\hb_i,z)$.

Note that $\W_\hb(\sw_2)$ is the VOA of the Virasoro algebra. Its embedding
into $\pi^\hb_0$ has been known for a long time \cite{ct}. In was used by
Feigin-Fuchs \cite{FeFu} to study representations of the Virasoro algebra,
and by Dotsenko-Fateev \cite{df} to obtain correlation functions of the
minimal models, in the same way as in \secref{sol}.

The VOA $\W_\hb(\sw_3)$ was first constructed by Zamolodchikov
\cite{zam}, and the VOA $\W_\hb(\sw_n), n>3$, was first constructed by
Fateev-Lukyanov \cite{fl} (cf. also \cite{bg}). The existence of
$\W_\hb(\g)$ as a VOA ``freely'' generated by currents of degrees $d_i+1$
for an arbitrary $\g$ was an open question until \cite{fre,ff:im}.

To summarize, we defined the VOA of a $\W$--algebra as a vertex operator
subalgebra of a VOA of free fields, subject to a set of constraints. The
constraints satisfy certain algebraic relations, which make it possible to
describe the structure of the $\W$--algebra: the operators $\q^\hb_i$
generate the nilpotent part of $U_q(\g)$, so that $U_q(\g)$ and
$\W_\hb(\g)$ form a ``dual pair''. The classical origin of these
constraints is a non-linear integrable equation: the Toda equation, and
therefore the classical limit of a $\W$--algebra consists of local
integrals of motion of this equation. In our forthcoming joint work with
Feigin we will derive the Wakimoto realization in a similar fashion.

\subsection{Quantum Drinfeld-Sokolov reduction.}    \label{ds}
$\W$--algebras can also be defined through the quantum Drinfeld-Sokolov
reduction \cite{ff:gd,fre}. Let ${\cal C}$ be the Clifford algebra with
generators $\psi_\al(n), \psi^*_\al(n), \al \in \De_+, n \in \Z$, and
anti-commutation relations $$[\psi_\al(n),\psi_\beta(m)]_+ =
[\psi^*_\al(n),\psi^*_\beta(m)]_+ = 0, \quad
[\psi_\al(n),\psi^*_\beta(m)]_+ = \delta_{\al,\beta} \delta_{n,-m}.$$
Denote by $\bigwedge$ its Fock representation, generated by vector $v$,
such that $\psi_\al(n) v = 0, n\geq 0, \psi^*_\al(n) v = 0, n>0$. This is
the super-VOA of ${\cal C}$. Introduce a $\Z$--grading on ${\cal C}$ and
$\bigwedge$ by putting $\deg \psi^*_\al(n) = - \deg \psi_\al(n) = 1, \deg v
= 0$.

Now consider the complex $(V_k \otimes \bigwedge,d)$, where $V_k$ is the
VOA of $\g$ of level $k$, and $d = d_{\on{st}} + \chi$. Here $d_{\on{st}}$
is the standard differential of semi-infinite cohomology of $\N_+$ with
coefficients in $V_k$ \cite{fei}, and $\chi = \sum_{i=1}^l
\psi^*_{\al_i}(1)$ corresponds to the Drinfeld-Sokolov character of $\N_+$
\cite{ds}. The cohomology $H^*_k(\g) = \oplus_{\n\in\Z} H^n_k(\g)$ of this
complex is a VOA \cite{ff:gd}. This cohomology can be computed using the
spectral sequence, in which the $0$th differential is $d_{\on{st}}$ and the
first differential is $\chi$.

\smallskip
\noindent{\bf Proposition 2} \cite{ff:gd,fre} {\em For generic $k \neq
-h^\vee$ the spectral sequence degenerates into the complex
$F_{1/(k+h^\vee)}^*(\g)$. Thus, $H^0_k(\g) \simeq \W_{1/(k+h^\vee)}(\g)$ and
$H^i_k(\g) = 0, i\neq 0$.}
\smallskip

The second part of Proposition 2 was proved for an arbitrary $k$ in
\cite{dt} using the opposite spectral sequence.

For any module $M$ from the category ${\cal O}$ of $\G$, the cohomology of
the complex $(M \otimes \bigwedge,d)$ is a module over the $\W$--algebra
$I_{1/(k+h^\vee)}(\g)$. This defines a functor, which was studied in
\cite{fkw}.

The limit of the $\W$--algebra $I_{1/(k+h^\vee)}(\g)$ when $k=-h^\vee$ is
isomorphic to the center $Z_{-h^\vee}(\G)_{\lo}$ of $U_{-h^\vee}(\G)_{\lo}$
\cite{ff:gd,fre}. It can also be identified with $I(\g^L)$, where $\g^L$ is
the Langlands dual Lie algebra to $\g$ \cite{ff:gd,fre}. This proves
Drinfeld's conjecture that $Z_{-h^\vee}(\G)_{\lo} \simeq I(\g^L)$, which
can be used in the study of geometric Langlands correspondence.

\subsection{Affine Toda field theories.} An analogue of the complex
$F_\hb^*(\g)$ can be constructed for an arbitrary Kac-Moody algebra. In the
case of an affine algebra its first cohomology can be identified with the
space of local integrals of motion of the corresponding affine Toda field
theory \cite{ff:im}.

Let us first consider the classical case \cite{ff:im}, \S~3,
\cite{ff:kdv}. The Toda equation associated to an affine algebra $\G$ is
given by formula
\eqref{toda}, in which the summation is over $i=0,\ldots,l$, and $\phi_0(t)
= -1/a_0 \sum_{i=1}^l a_i \phi_i(t)$, where $a_i$'s are the labels of the
Dynkin diagram of $\G$ \cite{K}; $\phi_0(t)$ corresponds to the extra
root $\al_0$ of $\G$. Following the scheme of \secref{to}, we define the
space $I(\G)$ of local integrals of motion as $$I(\G) = \bigcap_{i=0}^l
\on{Ker}_{\F_0} \Q_i.$$

Using the BGG resolution of the affine algebra $\G$ \cite{rw}, we can
construct a complex $F^*(\G)$ in the same way as in \secref{nil}. Now the
operators $\q_i: \pi_0 \arr \pi_{-\al_i}, i=0,\ldots,l$, generate the
nilpotent subalgebra $\widetilde{\n}_+$ of $\G$. The cohomology of the
complex $F^*(\G)$ coincides with the cohomology of $\widetilde{\n}_+$ with
coefficients in $\pi_0$, $H^*(\widetilde{\n}_+,\pi_0)$.

\smallskip
\noindent{\bf Theorem 7} \cite{ff:im,ff:kdv}
$H^*(\widetilde{\n}_+,\pi_0) \simeq \bigwedge^*(\goth{a}_+^*)$, {\em and}
$I(\G) \simeq H^1(\widetilde{\n}_+,\pi_0) = \goth{a}_+^*$, {\em where}
$\goth{a}_+^*$ {\em is the dual space to the principal abelian subalgebra
of} $\widetilde{\n}_+$.
\smallskip

Theorem 7 implies that local integrals of motion of the Toda theory
associated to $\G$ have degrees equal to the exponents of $\G$ modulo the
Coxeter number. The corresponding hamiltonian equations form the modified
KdV hierarchy \cite{ds,kw}. In \cite{ff:kdv} we gave a geometric
interpretation of these equations.

We can now define the space $I_\hb(\G)$ of quantum integrals of motion as
$$I_\hb(\G) = \bigcap_{i=0}^l \on{Ker}_{\F^\hb_0} \Q^\hb_i.$$ Using the BGG
resolution over $U_q(\G)$ we proved in \cite{ff:im} that all classical
integrals of motion can be quantized, so that $I_\hb(\G) \simeq
I(\G)$. Note that for $\G=\widehat{\sw}_2$ this was conjectured in
\cite{ger}. The quantum integrals of motion form an abelian subalgebra of
the $\W$--algebra $I_\hb(\g)$. They can be viewed as conservation laws of
integrable perturbations of conformal field theories associated to
$\W_\hb(\g)$ \cite{z,ey,hm}.

Free field realization, which at first appeared as a technical tool for
computing correlation functions of conformal field theories, has evolved in
the last few years to a powerful method of representation theory of
conformal algebras. There is every indication that similar ideas are
applicable to a much broader class of models of quantum field theory and
statistical mechanics, and to related representations.

\end{document}